\documentclass[prb,preprint,showpacs]{revtex4}

\newcommand{\dsi}{\vec{\delta S}_i(t)}

\newcommand{\dsja}{\delta S_j^{\alpha}(t)}
\newcommand{\si}{\vec{S}_i}
\newcommand{\ri}{\vec{R}_i}
\newcommand{\sj}{\vec{S}_j}
\newcommand{\B}{\vec{B}_{\rm ext}}
\newcommand{\gm}{g \mu_B}

\newcommand{\la}{\lambda}

\newcommand{\beq}{\begin{equation}}
\newcommand{\eeq}{\end{equation}}
\newcommand{\beqnar}{\begin{eqnarray}}
\newcommand{\eeqnar}{\end{eqnarray}}
\newcommand{\bfig}{\begin{figure}}
\newcommand{\efig}{\end{figure}}

\usepackage{graphicx}
\usepackage{graphics}  % to use for eps graphic file inclusion
\usepackage{dcolumn}
\begin{document}
\title{Spin dynamics characterization in magnetic dots}

\author{Mohammad-Reza Mozaffari, Keivan Esfarjani }
\email{k1@sharif.edu}
\affiliation{Department of Physics, Sharif University of Technology,
Tehran, 11365-9161 IRAN}

\date{\today}

\begin{abstract}
The spin structure in a magnetic dot, which is an example of a quantum few-body system, is studied as a function of exchange coupling strength and dot size within the semiclassical approximation on a discrete lattice. As the exchange coupling is decreased or the size is increased, the ground state undergoes a phase change from a single domain ferromagnet to a spin vortex. The line separating these two phases has been calculated numerically for small system sizes.
%, and analytically for larger dots. 
The dipolar interaction has been fully included in our calculations.
Magnon frequencies in such a dot have also been calculated in both phases by the linearized equation of motion method. These results have also been reproduced from the Fourier transform of the spin autocorrelation function.
From the magnon Density Of States (DOS), it is possible to identify the magnetic phase of the dot.
Furthermore, the magnon modes have been characterized for both the ferromagnetic and the vortex phase, 
and the magnon instability mechanism leading to the vortex-ferro transition has also been identified.
The results can also be used to compute finite temperature magnetization or vorticity of magnetic dots. 
\end{abstract}

%\pacs {{ 75.75.+a}{}}

\maketitle

\section{Introduction}

Developments in the nanomagnet fabrication technology have attracted much attention of
the physicists in the past decade\cite{experiment}. Nanomagnets can be used as memory elements\cite{cowburn}, magnetic field sensors\cite{ross}, computing and logic operation devices\cite{cowburn-device}. It is thus important to understand their static and dynamic behavior in both the single-domain ferromagnetic(SDF) and vortex phases, in thin film samples.
They are, furthermore, a good example of a few-body system to test models and theories used in micromagnetic calculations.
These so-called magnetic quantum dots are made of permalloy materials (Fe-Ni) deposited on a nonmagnetic semiconducting substrate such as Si. Their size ranges from ten to a few hundred nm, and their thickness is about 20 nm. For this reason, they are treated as two dimensional (2D) systems.
Due to dipolar interactions, a vortex phase can form in large enough ferromagnetic dots which are made of permalloy or supermalloy materials.
For fixed exchange coupling, a phase diagram for the stability of the vortex phase has been computed and compared to experiments as a function of dot size and thickness\cite{cowburn}. The results of this micromagnetic theory agrees relatively well with the experiments. Usov and Peschany have used a variational ansatz for the vortex arrangement of the spins in a disk-shaped dot, and studied its ground state structure\cite{variational}.
Also, Chui and Ryzhov\cite{chui} have used Monte Carlo and analytical methods in order to investigate the vortex state of a rectangular dot.
The effect of an in plane field, which is to move the vortex core, was studied analytically by Guslineko and Metlov\cite{guslienko-metlov}. 
Guslienko et al.\cite{guslienko-reversal} have used a micromagnetic model as well as a variational calculation\cite{variational} to compute 
reversal fields in a dot where the ground state is a vortex. They have also computed the hysteresis loop and have identified the modes causing the instability of the vortex phase: the so called C-shape and S-shape modes.
 
Since instabilities are of dynamical origin, and also because of the importance
of identifying the excitations in a magnetic system, it is very important to compute 
the magnon frequencies and characterize their oscillation modes in such dots. In this direction, the first steps 
were taken\cite{guslienko-thiele} by using Thiele's equation\cite{thiele}. This work
has studied the mode corresponding to the oscillations of the vortex core with 
an eventual damping (Landau-Lifshitz-Gilbert equation used within a micromagnetic solver).
Ivanov et al.\cite{ivanov} have computed analytically the few lowest magnon frequencies as a function of the dot radius. Dipolar interactions were replaced by imposing the boundary condition that the magnetization be tangent to the dot circumference. In addition to the oscillations of the center, they have identified the second mode as oscillations of the core size.  In another work, Guslienko et al. considered magnons in a square shape dot\cite{guslienko-square}. All these calculations were based on micromagnetic theory and 
continuum modeling. Furthermore, only the lowest modes were identified and calculated.

In this work, we have considered a discrete model of a magnetic dot, including explicitly the dipole interaction term. After identifying its different phases, the phase diagram in the (exchange coupling - dot size) plane is calculated. 
%we compute all the magnon modes and frequencies. A density of states is extracted for both single-domain and vortex phases.
%The calculations are also compared
%with the spectrum obtained from the Fourier transformation of the spin autocorrelation function.
%A renormalization scheme allows us to identify the exchange coupling and other parameters of our model Hamiltonian with those of a given experimental sample.
%The next section deals with the renormalization procedure and extraction of the model Hamiltonian parameters. 
Section \ref{magnons} treats the dynamics where 
the equations of motion are linearized. Magnon frequencies are obtained in two ways: diagonalization of the linearized equations of motion and Fourier transform of the spin autocorrelation function. 
Modes are then characterized for both the vortex and the ferromagnetic cases. The lowest modes, which are responsible for
the instability near the transition region have been identified. The paper is ended with conclusions.
 
\section{Method}

We consider a finite set of spins with exchange and dipolar interactions in an eventual magnetic field. We assume that there is no disorder present in the sample, the only source of anisotropy is magnetostatic. Magnetocrystalline anisotropy is neglected in this work as the dipole-induced shape anisotropy is enough to cause the spins to lie in the plane and make a vortex (if the exchange is small enough). 
As the relaxation time due to nonlinear magnon-magnon interactions or magnon-phonon coupling is usually of 
the order of nanoseconds and thus larger than typical magnon periods, the latter are well-defined 
excitations. Thus the inclusion of the Gilbert damping will only give them a finite lifetime and will not
affect the frequencies. For this reason, it is neglected in this work.

The Hamiltonian for this system can be written as follows:

\begin{widetext}
\beq
{\cal H}= -{1 \over 2} \sum_{<i\neq j>} J_{ij}\, \si . \sj 
+ \gm \sum_{i} \si . \B + {\mu_0 \over 8 \pi} (\gm)^2 \sum_{i \neq j}  {\si . \sj \over R_{ij}^3 }- {3(\si . \vec{R_{ij}}) (\sj . \vec{R_{ij}}) \over R_{ij}^5 }
\eeq
\end{widetext}
where $\B$ is the applied magnetic field, and $J_{ij}$ is the exchange integral between spins
$i$ and $j$. The latter is short-ranged and strong. Usually in permalloy systems the 
exchange integral is of the order of a few tenths of an eV. The dipolar interaction, however, is much weaker, by
three orders of magnitude but is long-ranged. Therefore it becomes important in larger
samples, and needs to be taken into account. It can furthermore account for the anisotropy in the sample. 
Unlike most calculations where the demagnetization field is included as a boundary condition, we explicitly include the dipole interaction in our calculations as indicated in the above Hamiltonian.

In this paragraph, we discuss in qualitative terms the physics and behavior of a vortex phase under magnetic fields.
For small samples the exchange term ($J>0$) dominates and the ground state is a single domain ferromagnet. 
For large enough samples, or small enough exchange coupling, the magnetostatic (dipolar) energy term becomes 
dominant, and the ground state of a disk becomes a vortex. 
In this case, no lines of field leak outside the sample and thus the magnetostatic energy, which usually has a large contribution in the total energy, becomes minimum. 
If the film thickness becomes comparable or larger than the disk radius, then a ferromagnetic state develops in the core of the cylinder. Indeed in the core region, the vortex configuration is unfavorable compared to an exchange energy driven ferromagnetic configuration, and thus the spins
tend to have a slight inclination along the axis perpendicular to the disk. For thick enough disks, the core region can
develop a magnetization parallel to the disk axis. The core radius is thus an increasing function of the thickness\cite{variational}.
An in-plane external field will shift the center of the vortex away from the center, so as to make the regions of magnetization parallel to the field larger. At larger fields, the vortex core is  repelled out of the sample 
area and the system becomes fully ferromagnetic\cite{metlov,variational}. 
A hysteresis curve can be obtained for the vorticity and the magnetization as a function of the external field. 
A field perpendicular to the plane of the disk either creates a ferromagnetic core if the 
latter does not exist, or will widen the core radius if it already exists.

In this work, we intend to compute the ground state and the magnetic excitations of quantum dots for both the ferromagnetic and the vortex states of a monolayer dot in the absence of an external field, by using a semiclassical approximation. 

\subsection{Ground state calculations}

The magnons are the spin excitations above the ground state. It is therefore necessary to find first the ground state of this Hamiltonian. This can be achieved by minimizing, within the mean-field approximation, the total energy with respect to the spin configuration. As a result, one finds that each spin is aligned along the molecular field at its site. The latter is given by:
\begin{widetext}
\beq
\vec{B}_{\rm eff}(i) ={1 \over \gm}{\partial{\cal H} \over \partial \si} =\vec{B}_{\rm ext} - \sum_{j\neq i}  {J_{ij} \over \gm} \sj + {\mu_0 \over 4 \pi} \gm \sum_{j\neq i}  {\sj \over R_{ij}^3 }- {3 \vec{R_{ij}} (\sj . \vec{R_{ij}}) \over R_{ij}^5 }
\eeq
\end{widetext}
If a good starting configuration is guessed, then one can simply iterate, with an
eventual mixing scheme, the Euler-Lagrange equations which simply state that each spin 
must lie along the effective (Weiss-)field. This is obtained from the minimization of the total energy with the constraint of spin normalization. In case no good starting guess 
is known, one can perform a Monte-Carlo Simulation at a finite temperature and anneal
the system to reach the ground state. Typically a spin is picked at random, rotated
at random, then the total energy change of the system is computed and compared to $k_BT$.
The move is accepted if $e^{-\Delta E/k_BT}>r$ where $r$ is a random number in [0,1];
otherwise, the original configuration and total energy is kept and another spin is chosen and rotated at random. This is called the Metropolis algorithm\cite{metropolis}. Depending on 
the size of the system, one needs to perform many moves in order to reach equilibrium
at temperature T. Lowering the temperature slowly enough guarantees that the true ground state
will be reached at the end of the simulation.

\subsubsection{Energetics and Phase diagram}

In this section, we give the expressions for different terms in the total energy, and discuss the phase stability. 
Starting from a large value for the exchange integral $J$, and a fixed lattice size,
one can calculate the ground state and lower $J$ to investigate the phase change.
We will only consider a two-dimensional disk-shape geometry. It is well-known that for high enough $J$ the ground state is ferromagnetic. As $J$ is decreased, the system goes
through a phase change: the ground state becomes a single vortex with its core localized at the center of the dot\cite{variational,metlov}. 
As $J$ is further decreased, we have discovered that
there are more phase changes, the ground state may have a higher number of vortices actually generated from higher magnon modes (to be discussed in the section on magnons). For $J=0$ it was found that the ground state can be seen as 
%a bent line of dipoles or alternatively, as 
a "crystal" of smallest possible vortices sitting near each other and forming a vortex lattice.
In what follows, we will be interested in dots with one vortex at the most, 
i.e. the exchange integral $J$ does not become too small, and
we will only be interested in the single-ferromagnetic-domain - vortex (SFD-V) transition. 
In both phases, the exchange energy is mainly proportional to $N J S^2$. The difference between vortex and the single domain exchange energy comes from the sum of the core part which is $E_{\rm core}=e_{\rm core} J S^2$, independent of $N$, and the long-range logarithmic term characteristic of vortices, proportional to Log $N$.
The core energy in units of $J S^2$ can be deduced to be : $e_{\rm core}=2.29$ from a fit to 
numerical data for a square lattice and a disk-shaped dot.
Therefore the difference between the exchange energy of the vortex and the SDF phase is equal to : $JS^2\, (e_{\rm core} + {\pi \over 2} {\rm Log} N)$.

On the other hand, the discretized dipole energy can be approximated in the $a \to 0$ limit ($a$ is the lattice 
constant), as the continuum approximation $E^{\rm dipole}=- (\mu_0/2 )\,\int M.H d^2r$ plus a self energy 
correction due to diagonal $i=j$ terms. The field $H$ is the macroscopic field satisfying $B=\mu_0(H+M)$.
The self energy correction is thus extensive, and can be 
written as $E^{\rm self-energy}=( \alpha \,N + \beta \sqrt{N})\, \mu_0 \,(\gm S)^2 /a^3 $. The
first term, $\alpha$, coming from the bulk contribution, and the second, $\beta$, from boundary atoms. 

%These two terms can be obtained from a fit to numerical data for the magnetostatic energy, since the continuum formula is easily calculable for both phases as we will see below.

In the {\bf fully ferromagnetic} phase, where all spins have the same exact direction, the continuum form of the 
dipole energy, after assuming $H=-\gamma \, M_{\rm 2d}/a$ , is reduced to 
\begin{widetext}
$$ 
E_{\rm SDF}^{\rm dipole} =  
(\gamma \mu_0 /2a) \,\int M^2 \, d^2r = (\gamma \mu_0 /2a) (\gm S/a^2)^2 (N a^2) = N \,{\gamma \over 2} \times  \mu_0  (\gm S)^2/ a^3  
$$ 
\end{widetext}
Here $\,\gamma$ is the demagnetization factor and $M$ or $M_{\rm 2d}$ the two-dimensional magnetization.
In the vortex phase, however, the continuum limit of this energy reduces to zero as there are no magnetic 
charges: we can write $H=\nabla \phi$ and $E^{\rm dipole}_{\rm V} \propto \oint \phi M.n \,dl - \int \phi \nabla .M d^2r = 0$ as the magnetization field in a vortex is divergenceless and tangent to the dot boundary.

To summarize, the total energy (exchange plus dipole) of the {\bf ideal} SDF and vortex dots can be written as:
\begin{widetext}
\beqnar
  E_{\rm SDF}& \simeq & -{1 \over 2} J S^2 N (z-{A \over \sqrt N}) +   \,{\mu_0 (\gm S)^2 \over a^3} \, N 
(\,\alpha + {\beta \over \sqrt{N}}) + \,{\gamma \over 2} \,{\mu_0 (\gm S)^2 \over a^3} \, N \\
  E_{\rm V}& \simeq & -{1 \over 2} J S^2 N (z-{A \over \sqrt N}) + 
\,{\mu_0 (\gm S)^2 \over a^3}\, N (\,\alpha + {\beta \over \sqrt{N}}) +
(E_{\rm core} + {\pi J S^2 \over 2} {\rm Log} N)    \\
\eeqnar
\end{widetext}
In the above, $z$ is the number of nearest neighbors; the second term $-A/\sqrt N$ is added in order to 
include boundary atoms which experience a different environment). For a disk-shaped dot forming a square 
lattice, $z=4$ and $A=4.52368 $, and the demagnetization factor is $\gamma=1/2$ for a perfectly ferromagnetic 
sample with circular geometry. 
%We have found however, that the above formula can not be used in order to obtain characteristics of the critical point, as near the transition to the vortex phase, $J$ is weak and the spin configuration of the sample is not fully parallel. Therefore the formula $H=-\gamma \, M_{\rm 2d}/a$ is not valid anymore.
Furthermore, from a fit to dipole energies of this same dot, we obtain $\alpha=-0.188$ and $\beta=0.0925$.
where the constant ${\mu_0 (\gm S)^2 \over a^3}$ is equal to 0.3396 meV for $a=2\AA, \, g=2$ and $S=1$.

% -2.224 N+11.38 \sqrt{N} vortex
% -2.2277 N+7.488 \sqrt{N} ferro

%From the above formulas, it can be deduced that, for a given size $N$, the critical exchange coupling is determined from the competition between the core and logarithmic energy of the vortex and the demagnetization energy of the SDF:
%\beq
%J_{\rm critical} = {\mu_0 (\gm )^2 \over a^3} {{1 \over 4} N \over e_{\rm core}+ {\pi \over 2} \,{\rm Log}\, N};\,\,
%\label{jcritical}
%\eeq
 
This analysis is valid for ideal vortices and SDF samples where the 
magnetization direction is respectively circular and straight. For finite size 
samples near the critical point, however, there will be a slight deviation 
from the ideal orientation and the formula $H=-\gamma \, M_{\rm 2d}/a$ is not 
valid in the SDF case anymore. 
Furthermore, the dipole energy of the vortex will not be exactly equal to zero.
For this reason, the phase boundary is computed numerically.

We have performed relaxation calculation for finite size dots. The total energy calculations were done for a square and a circular shape dot of thickness one (a monolayer). The critical exchange coupling parameter was obtained and plotted as a function of the dot size (number of sites). %The results are displayed in Fig. \ref{phase} .
In the calculations, the spin magnitude S was taken to be 1; the lattice constant was $a=2\AA$, and the lattice was of square type. The two curves were fitted with $J_c(eV) = 1.436\,\times 10^{-5} \,N^{0.4843}$ for the circle, and $J_c(eV) = 1.672 \,\times 10^{-5} \, N^{0.4147}$ for the square. These graphs are displayed in  Fig. \ref{phase}.
It seems that for circular dots, %if the critical line is computed without relaxing the spins, 
the line $J_c = 1.28 \,\times 10^{-5} \, \sqrt{N}$ is a good fit and is also plotted on the graph.
We have not found however a simple explanation for this size-dependence of $J_c$.
From this study, it can be concluded that at or near the critical point ($J \propto \sqrt N$), even though dipole and i
exchange fields are of the same order, the dominant term in the total energy is the exchange term as it 
behaves nearly like $N^{3/2}$ whereas the dipole energy is at most linear in $N$.
Furthermore, as the two phases are separated by a non zero potential
barrier for $J\simeq J_c$,  the transition is similar to a ``first order" one  defined only for an infinite system.
\bfig[h]  
\begin{center}
\includegraphics[angle=270, width=7 cm]{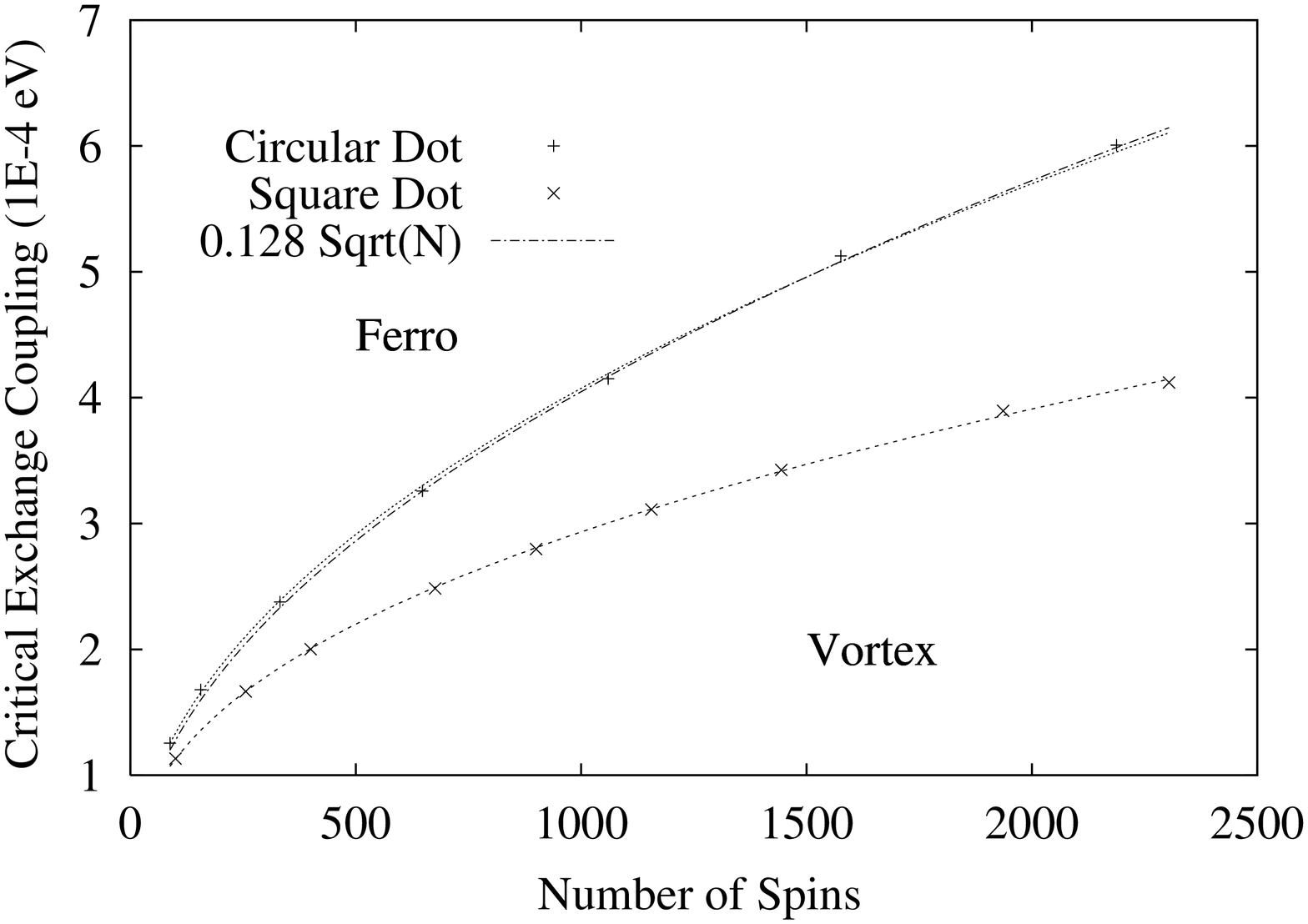}
\end{center}
\caption{Phase diagram of a square and circular dot as a function of size and exchange coupling. The lines are the fitted curves given in the text.} \label{phase}
\efig 

Finally, the parameters which appear in the energy functions of the two idealized
phases are summarized in table \ref{params}.
\bigskip

\begin{table}[h]
\begin{tabular}{|c|c|c|c|c|}

\hline

 $\alpha$ & $\beta$ & $A_{\rm square} $ & $e_{\rm core}$ & ${\mu_0 (\gm )^2 / a^3}$ \\

\hline

 -0.188 & 0.0925 & 4.524 & 2.29 &  0.3396 meV \\

\hline

\end{tabular}
\caption{Numerical values of the parameters in the energy function}
\label{params}
\end{table}

\subsection{Spin dynamics and magnon calculations}

Lowest frequency magnon modes have been calculated in the continuum approximation\cite{ivanov} 
and also characterized experimentally\cite{park}. Below, we will use the semiclassical approximation 
(assuming each spin to be a classical dipole), and a discretized system of finite spins interacting 
via exchange and dipole fields, in order to compute magnon frequencies and characterize their modes.

Once the ground state spin configuration $\{\si^0\}$ is calculated from the 
mean-field equations, or the Monte Carlo algorithm, one can proceed to calculate small spin oscillations about this equilibrium: $\si(t)=\si^0 + \dsi$.
Assuming a harmonic dependence in time, and inserting this into the semiclassical equations of motion
\beq
{d\si \over dt} =  {\gm \vec{B}_{\rm eff}(i) \over \hbar} \times \si
\label{dynamics}
\eeq
and eliminating the term $\si^0 \times \vec{B}_{\rm eff}^0(i)=0 $, 
one obtains a system of Ricatti non-linear equations on $\dsi$.
Linearizing the latter with respect to $\dsi$, an
eigenvalue equation defining the magnon modes and frequencies will be obtained.
The effective field on site $i$ involves, through the exchange and magnetostatic interactions, the spin at other sites.
This makes the set of equations (\ref{dynamics}) a coupled set, which is given below:
\begin{widetext}
\beq
{d\dsi \over dt} =  {\gm \vec{B}_{\rm eff}^0(i) \over \hbar} \times \dsi +
\sum_{j \alpha} {\gm \over \hbar}{d \vec{B}_{\rm eff}(i) \over d S_j^{\alpha}}  \dsja \times \si^0
\label{dyn}
\eeq
\end{widetext}
Note that taking the dot product of the right side with $\si^0$ yields zero, implying that
the projection of the vector $\dsi$ on $\si^0$ does not change with time.
Thus the magnetization vector performs Larmor-like precessions around its ground state 
(equilibrium) value. % (or the mean field $\vec{B}_{\rm eff}^0(i)$).
One can write $\dsi_{\lambda}=(\alpha_{\lambda} \overrightarrow{u_i} +\beta_{\lambda} \overrightarrow{v_i}) 
e^{i \omega_{\lambda} t}$ for the magnon mode $\lambda$ and substitute it in Eq.(\ref{dyn}). 
The unit vectors $(\overrightarrow{u_i},\overrightarrow{v_i})$ are orthogonal to $\si^0$. 
This results in an eigenvalue equation whose solutions $\omega_{\lambda}$ are the magnon frequencies.
The type of oscillation about the ground state for that mode is characterized by the corresponding eigenvector 
defined by $(\alpha_{\lambda} \overrightarrow{u_i} +\beta_{\lambda} \overrightarrow{v_i})$.
The results on magnon frequency distribution and modes will be discussed in the next two sections.

\subsubsection{Magnon frequency distribution}

We have considered a 96 spin lattice in both vortex and SDF states.
The obtained magnon frequencies for each phase are displayed in Fig. \ref{magnons} for $J=0.6$ meV (SDF) and $0.1 $ meV (vortex)  respectively. 
We can observe a gap in the vortex spectrum whereas the spectrum of the SDF phase 
starts from zero frequency. 
Note that the gap (in units of $JS$) in the vortex phase will diminish as $J$ is increased. The lowest frequency mode in the SDF phase corresponds to in-plane collective oscillations or uniform rotations of the spins, if the weak anisotropy is neglected. The lowest frequency ($\omega \approx 0$), which is the Goldstone mode, will shift to a small non zero value if a weak in plane anisotropy is introduced. 
This occurs in larger samples. 
In our actual samples, this weak anisotropy exists due to the discrete nature of the 
lattice, and the lowest frequency is very small but non zero.

\bfig[h]
\includegraphics[angle=270, width=7 cm]{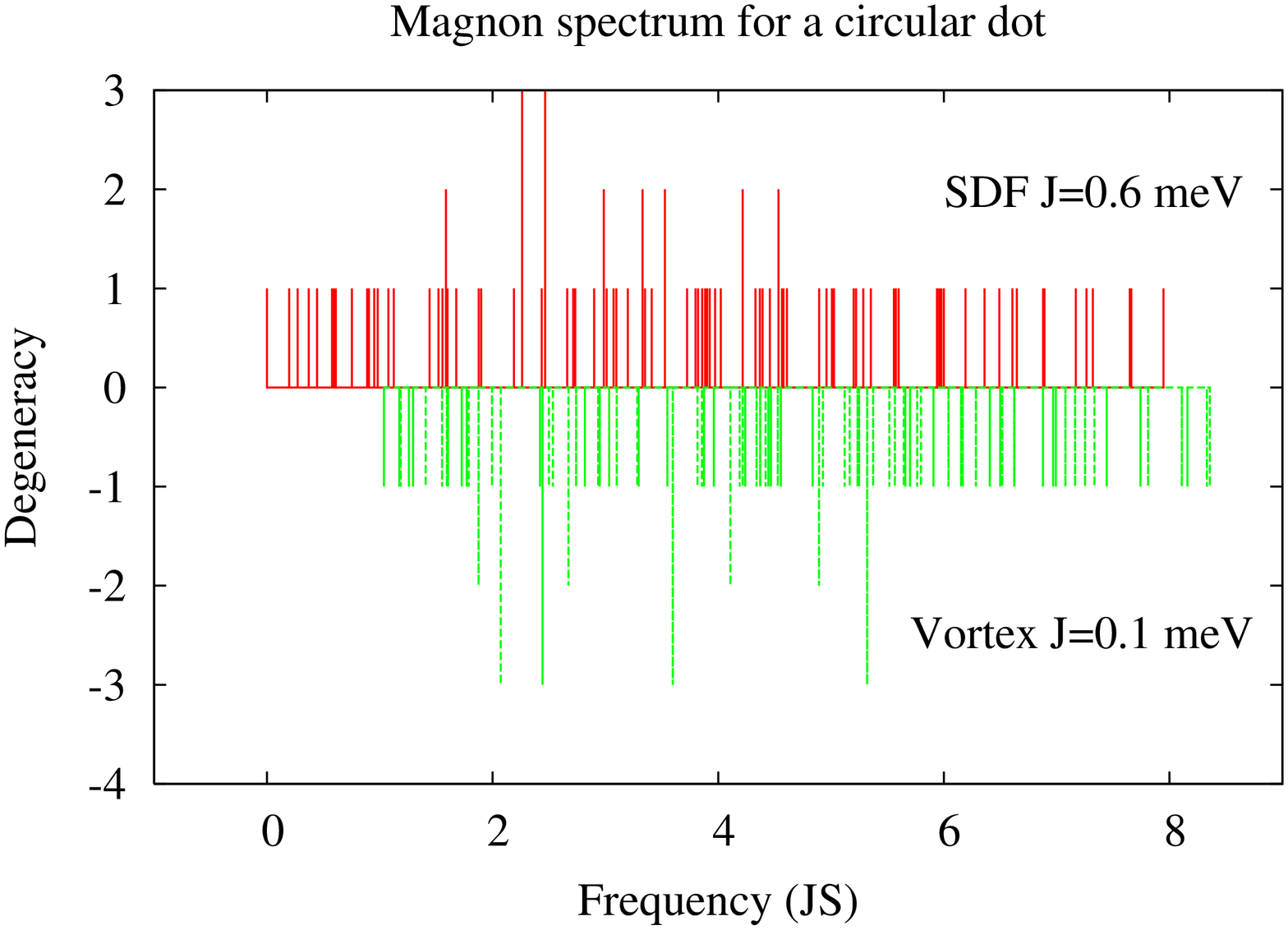}
\caption{Magnon spectra of a 96 spin dot with $J=0.6$ meV (ferromagnetic) and with
$J=0.1 $ meV (vortex) above and below the x axis respectively. To have them on the same 
scale, frequencies are plotted in units of $JS$. } 
\label{magnons}
\efig 

To check the correctness of the results, we have also performed an independent calculation of the magnon spectrum from the Fourier transform of the spin autocorrelation function defined as:
\beq
F(\tau)= {1 \over N T} \int_0^T  \sum_{i=1}^N \,
(\si(t)-\si^0) . (\si(t+\tau)-\si^0)^* \, dt
\label{autocorrelation}
\eeq
Here, the ensemble average has been replaced by the time average in which $T$ is a time scale larger than the largest magnon period so that all modes are sampled in this integral average.
Writing the spin at site $i$ and time $t$ as a general superposition of the eigenmodes $\vec{e}_{i \la}$:
$$\si(t) = \si^0 + \sum_{\la} \, (\alpha_{\lambda} \overrightarrow{u_i} +\beta_{\lambda} \overrightarrow{v_i})\, e^{i \omega_{\la} t} $$
substituting in equation (\ref{autocorrelation}), and using the orthogonality of the eigenvectors, one can easily show 
that its Fourier transform is of the form:
\beq
F(\omega)  = \sum_{\la}  \delta(\omega - \omega_{\la}) \, (|\alpha_{\lambda}|^2 +| \beta_{\lambda}|^2) 
\label{dos}
\eeq
which has peaks at precisely the magnon frequencies. 
The spin autocorrelation function was calculated by performing a spin dynamics simulation. 
The simulation was started with an arbitrary initial configuration near the true ground 
state. The spin trajectory at later 
times was then obtained by integrating equation (\ref{dynamics}) by using the finite difference method. 
Knowing the trajectories of all spins $\si(t)$ for a long enough time period, 
the calculation of the spin 
autocorrelation function is just a matter of summation and Fourier transformation. 
The spectra obtained by using this method, which in principle includes the nonlinear
deviations as well, are illustrated in Fig. \ref{spectra} and compared to the harmonic 
(analytical) results obtained by solving the linearized eigenvalue equation. 
As can be seen, the agreement 
is perfect provided the initial displacements are small enough. 
Even some of the doubly degenerate states are resolved in the nonlinear method.
%is very good confirming the correctness of both methods. 
The height of the peaks obtained from this method is given by the last term in Eq. 
(\ref{dos})  and is proportional to the amount of those modes present in the original 
spin configuration.

\bfig[h]
\includegraphics[angle=270, width=7 cm]{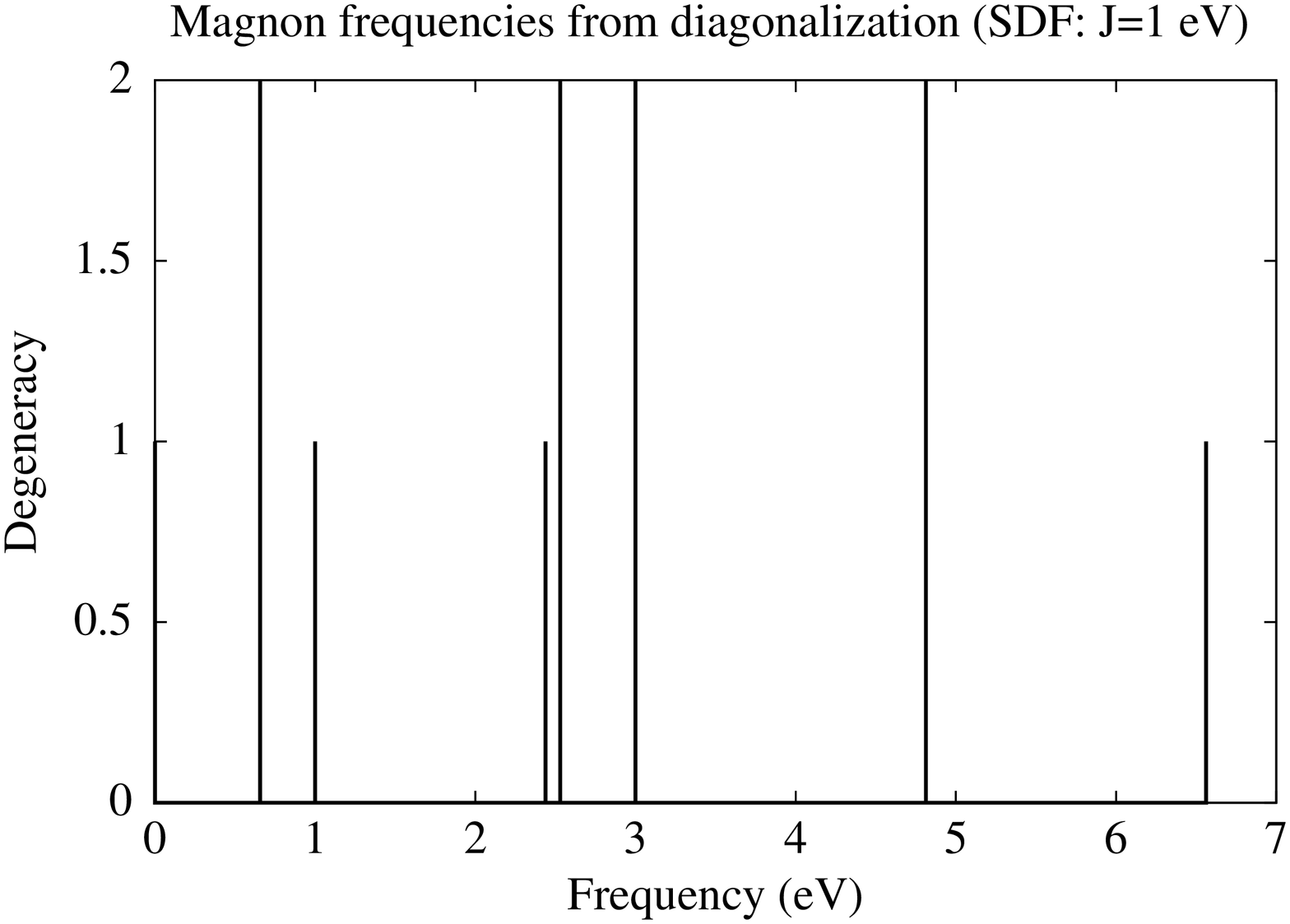} 
\nobreak
\includegraphics[angle=270, width=7 cm]{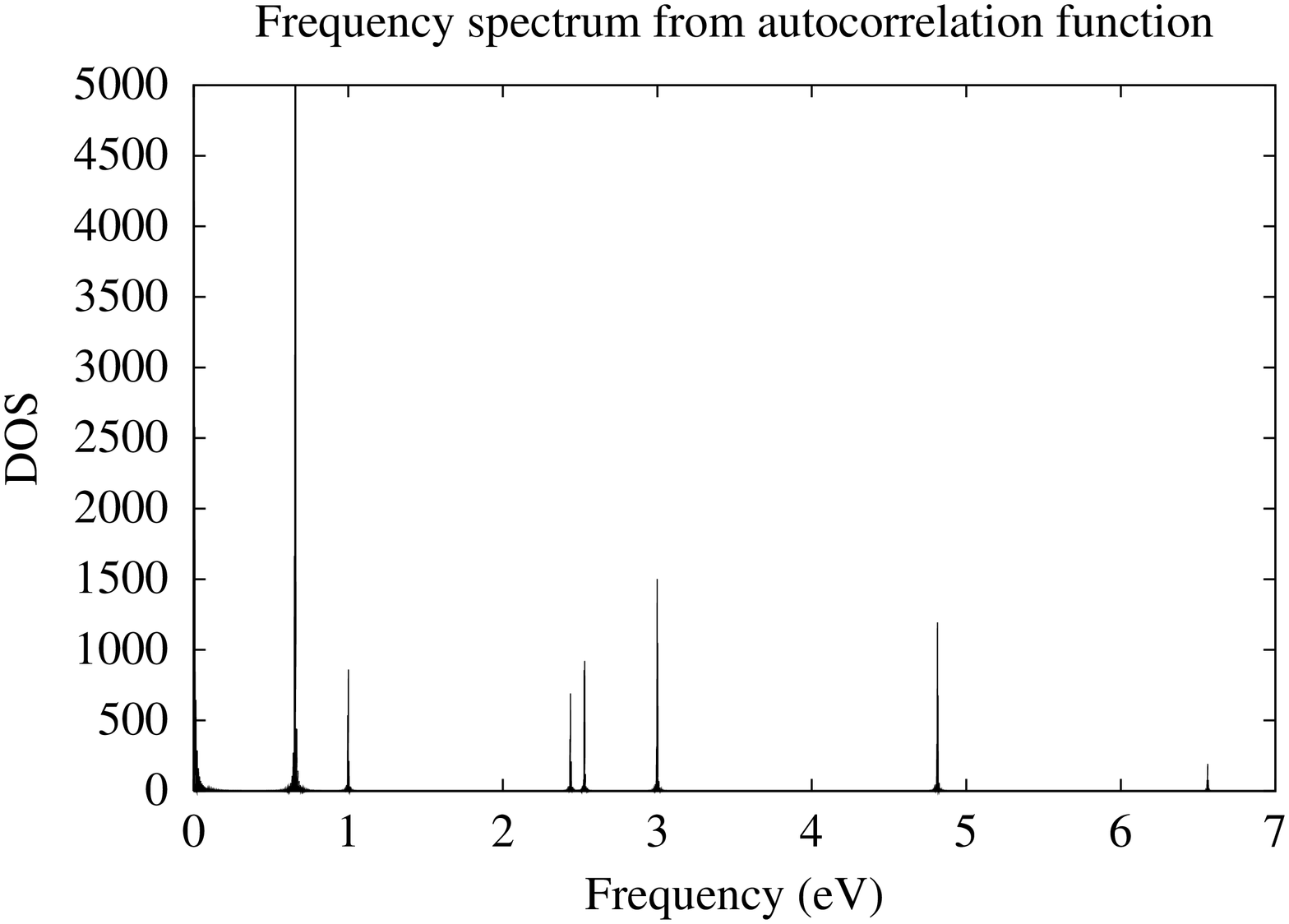}
\caption{Magnon spectra of a 12 spin dot with $J=1$ eV (ferromagnetic) obtained by diagonalization (left) and Fourier transform of the spin autocorrelation function (right).}
\label{spectra}
\efig 

The density of states (DOS) can also be deduced from our data. We have plotted in 
Fig. \ref{alldos} the DOS per spin in the SDF and the vortex phases of a circular sample 
for different dot sizes. For the considered sizes, $J=0.1 meV$ corresponds to a stable
vortex and $J=0.6 meV$ to a stable ferromagnetic phase. The DOS is defined as 
$$ DOS(\omega)  = \sum_{\la}  \delta(\omega - \omega_{\la}) $$
where, for practical purposes, the Dirac function $\delta$ was replaced by a broadened Gaussian. It can be seen that the bulk limit is reached for $N$ larger than a few thousand spins. 
The flat behavior at the band edges is characteristic of bulk 2D bands with quadratic dispersion.
\bfig[h]
\includegraphics[angle=270, width=7 cm]{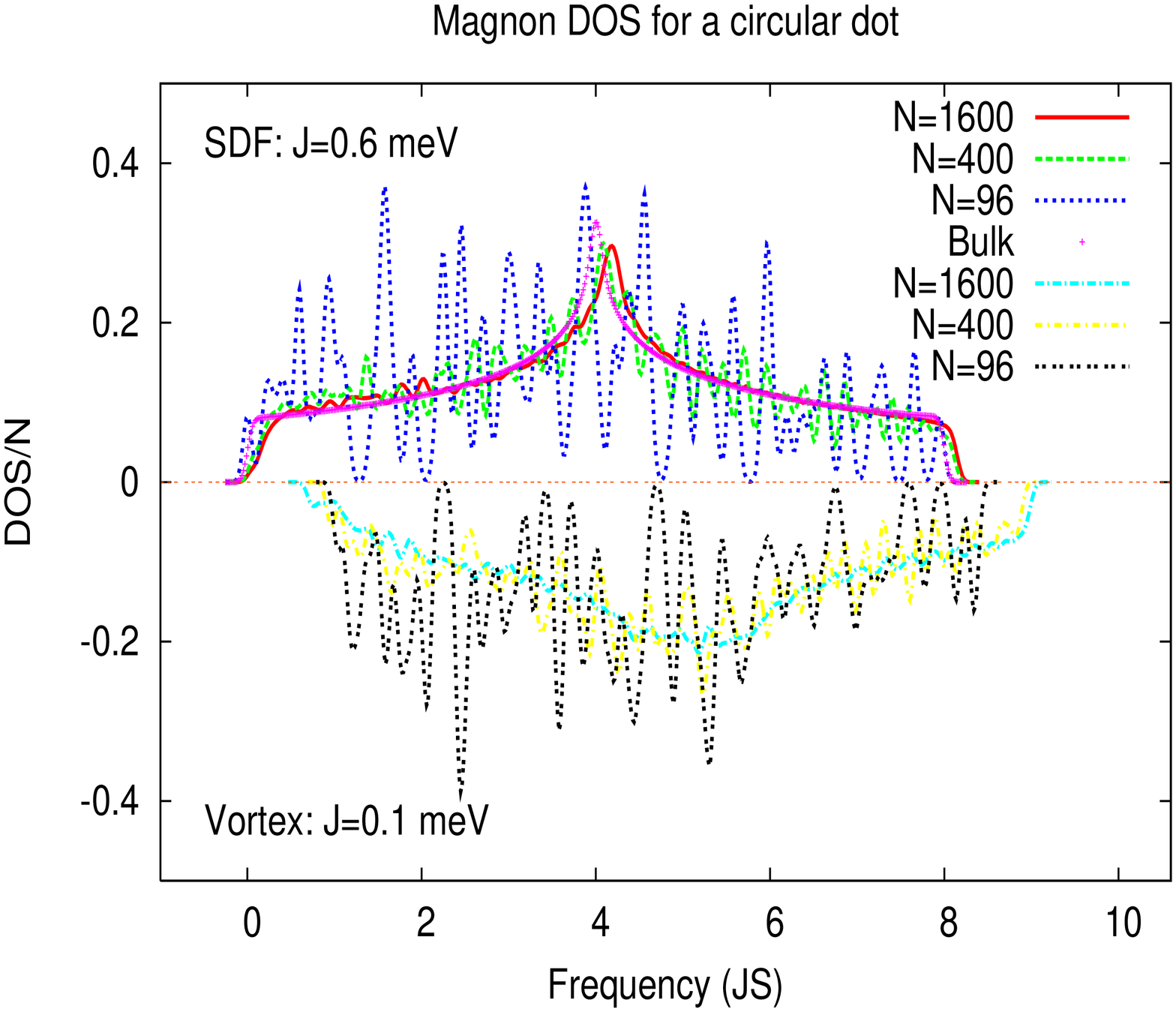}
\caption{DOS per atom in the SDF and vortex phases for 3 different dot sizes. Above the
axis data is obtained with $J=0.6 meV$ and below the axis data with $J=0.1 meV$. )}
\label{alldos}
\efig 

One can notice that the shapes in both phases are similar except for a relatively larger
gap of the vortex phase. Even for the same exchange coupling, frequencies of the vortex
phase are slightly above those of the SDF (see also Fig. \ref{instability}) 
The overall shape of the DOS is characteristic of 2D systems with a quadratic dispersion.
The DOS for a square lattice within the tight-binding model is also the same, 
namely it consists in a central peak separating two plateau-like regions.
The difference between the two phases resides in larger fluctuations and broadening 
at the band edges and center for the vortex phase. 
Due to broken symmetry, the vortex phase has usually a gap, whereas the SDF phase 
has a very low frequency Goldstone mode, the frequency  of which may go to zero 
for small enough $J$ or large enough sample size.
The latter can also be shifted to a nonzero value if additional anisotropy is 
present in the sample.
%\bfig[h]
%\includegraphics[angle=270, width=7 cm]{magnon-circle.ps}
%\caption{DOS per atom in the vortex phase for 3 different dot sizes}
%\label{vortexdos}
%\efig 

A similar magnon calculation for the vortex was also performed by Ivanov\cite{ivanov} using the continuum version
of the spin Hamiltonian. He could extract the lowest modes as a function of the dot radius and 
represented them in 2D with two quantum numbers (n,m) representing the number of nodes of the radial 
Bessel function and the index of the latter, respectively. However, only the lowest eigenmodes 
were discussed in their paper. They were identified as oscillations of the vortex position (m=1,n=0) 
and oscillations of the vortex core shape (m=0,n=0).
In this paper, all the modes are calculated and characterized. 
We found that the vortex phase
has yet another soft mode causing some instability which may eventually 
lead to the SDF phase. This and other modes will be discussed shortly.

\subsection{Temperature dependence of the Magnetization and Vorticity}

From these spectra, it is also possible to obtain the low-temperature dependence 
of the order parameter (magnetization or vorticity, see also Eq. \ref{op} for their definition) of the quantum dot: 
$$<S(T)>/S=1- \sum_{\la} n_{\rm BE}(\omega_{\la})/N_{\lambda}$$, where $ n_{\rm BE}$ is
the Boson distribution function and $N_{\lambda}$ is the total number of modes. 
In large enough ferromagnetic samples, one can assume $DOS(E)=\Theta(E) 
(D_0 + \alpha (E/JS)^2)$ at low energies, and 
will obtain the following i{\it low temperature} expansion: 
$$<S(T)>/S=1 - D_0 k_BT Log E_{\rm min}/k_B T - c \alpha (k_B T)^3 +... $$
where the positive constant $c$ is given by: $\,c=\int_0^{\infty} x^2 n_{\rm BE}(x) dx$ and $E_{\rm min}/$ is the
low energy cutoff of the spectrum, due to some kind of in-plane anisotropy,
leading to a tiny gap in the magnon spectrum.
For the vortex, however, there are two differences compared to the DOS of the ferromagnetic sample.
One is the presence of a gap, and two is the more smooth than a step function start of the DOS.
Assuming this start to be of the form $DOS(E)=E^s \Theta(E-E_{\rm min}); \,\,0<s<1$, 
one can analytically show that 
this results in a flat vorticity versus temperature until the latter reaches the gap value:
$ <S(T)>/S=1 - E_{\rm min}^{1+s} e^{-E_{\rm min}/k_BT} + ...$.
For larger temperatures, there is a small linear decrease of the vorticity.
We are not sure whether an experimental measurement of the vorticity versus temperature is
possible, but we predict that in the vortex phase the order parameter is constant as
temperature is increased from 0 until $k_BT$ reaches the value of the gap where it
starts to decrease almost linearly. As for the SDF samples, where the magnetization 
can be measured, we have 
predicted a superlinear decrease of the latter versus the temperature. 
The 2D system being finite, and anisotropy present, there is always 
a gap in the excitations and magnetic order is robust against small thermal fluctuations. 

%The calculations were performed for both vortex and SDF phases in samples 
%of size 100 and 1600 spins.
%Similar to the previous data on the DOS, we have considered a circular sample for the vortex phase ($J \lesssim J_c$)
%and a square sample for the SDF phase ($J \gtrsim J_c)$. As the samples are two dimensional, one would expect a 
%linear decrease of the order parameter as the temperature is increased. As is shown in Figs. \ref{moft} and \ref{voft}
%the SDF phase shows such behavior, but in the vortex phase, the linear decrease starts for temperatures higher than the gap in the excitations. It is therefore possible in principle, by measuring the temperature dependence of the magnetization or vorticity, to find out the nature
%of the ground state.

%\bfig[h]
%\includegraphics[angle=270, width=7 cm]{voft.ps}
%\caption{Normalized vorticity in the vortex phase for 2 different dot sizes of (N=96 and 1600)}
%\label{voft}
%\efig 
%\bfig[h]
%\includegraphics[angle=270, width=7 cm]{mag-vs-T.ps}
%\caption{Normalized magnetization in the SDF phase for 2 different square dot sizes (N=100 and 1600) calculated for 
%two different $J$ values near the critical point (J=0.4 meV) and away from it (J=0.1 eV).}
%\label{moft}
%\efig 

In the following section, we will discuss the obtained magnon modes.

\subsection{Magnon modes characterization}

The modes are the coherent libration of the spins on each site. The oscillations take place with the period 
associated with the frequency of that mode. One way to characterize them is by defining the nodal lines (in 2D).
The latter are the set of points at which the amplitude of the spin oscillations is zero (immobile spins).
This is very similar to the nodes in the eigenfunctions of an electron Hamiltonian. The eigenfunctions are
identified with their number of nodes: in 1D, the eigenstate number $n$ (if they are discrete) has $n-1$ nodes
along the x axis, excluding the node at infinity. So wavefunctions with a higher number of nodes in 1D, or 
number of nodal surfaces in higher dimensions, have a higher energy.
In the following, we will also use this concept in order to classify the magnon modes.

\subsubsection{SDF modes:}

In the ferromagnetic phase, they are well-known plane-wave type modes with half-wavelengths multiples of the dot 
size, but slightly deformed due to boundary effects. Dipole-induced anisotropy favors in plane precessions, 
but higher frequency modes having out of plane precessions also exist. The out of plane precession can take place 
both at the center, or at the boundary of the dot.
The lowest magnon mode is the in-plane and in-phase oscillations of the whole magnetization (Goldstone mode). 
The frequency associated with it is zero\cite{goldstone} or very small.
The second lowest one is a C-shape mode (see Fig. \ref{c-mode}) which consists in the bending of the magnetization
with a nodal line cutting the length of C into two. %A nodal line is defined as the location of spins which do not move.
Half of its wavelength is equal to the dot length and its wave vector is along the magnetization 
direction. The next mode is the S-shaped mode with a wavelength equal to the dot size. This mode posesses two nodal lines
dividing the length of S into three. A yet higher energy mode has one nodal line perpendicular to the magnetization,
or two nodal lines one along $M$ and the other perpendicular to it, so that
only spins at the four corners of the dot oscillate. Higher modes involve mostly motion of the outer spins, and 
have more nodal lines in both directions and of wavevectors of larger magnitude up to $\pi/a$.
In the highest mode, all spins in the central region precess in opposite phase to their neigbors, and the spins
at the boundary are immoblie due to confinement effects. For the next highest mode, this large-amplitude 
opposite-phase oscillations take place in two halves of the sample (a high energy p-wave state).
\bfig[h]
\includegraphics[width=7 cm]{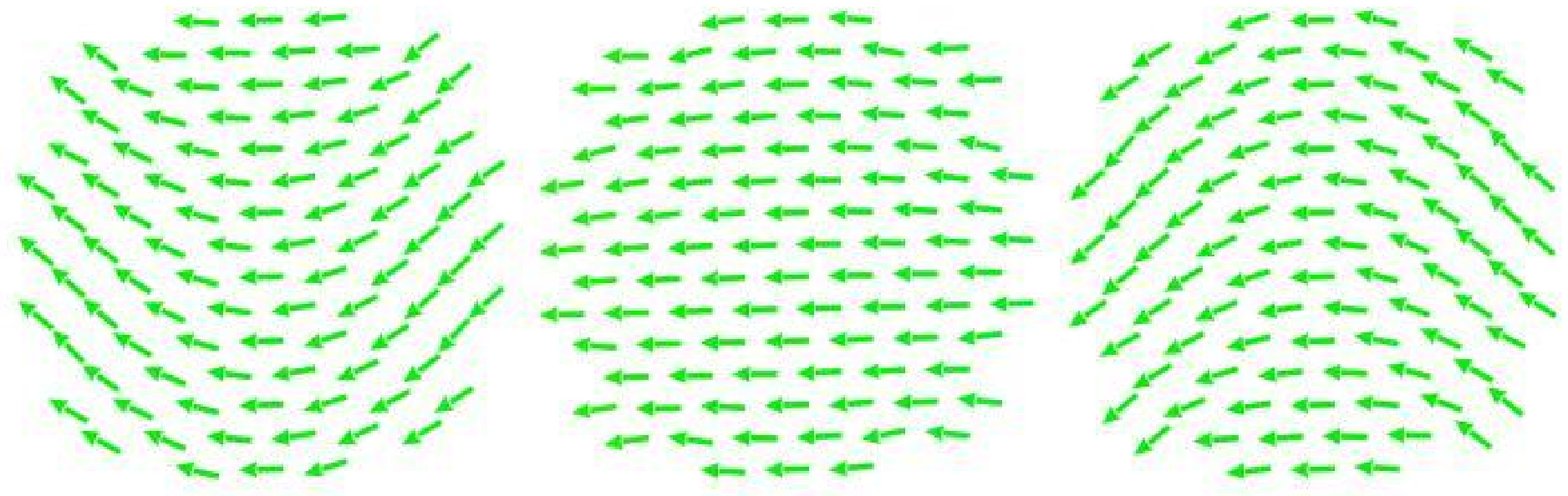}
\caption{C-mode in the SDF phase}
\label{c-mode}
\efig 
\bfig[h]
\includegraphics[width=7 cm]{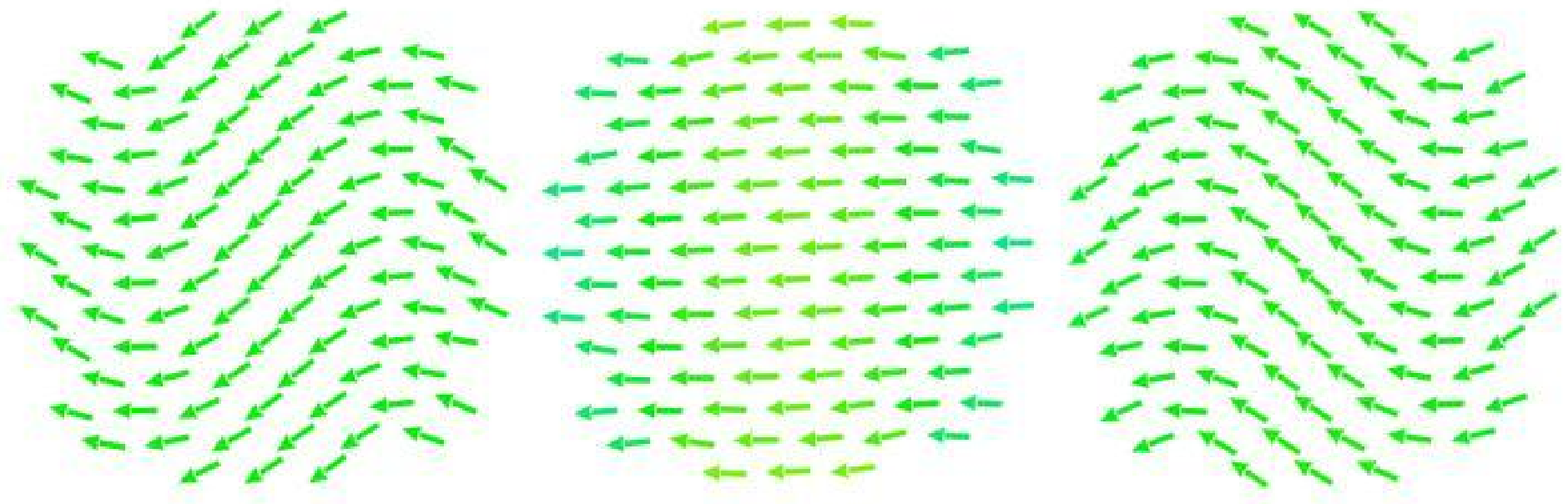}
\caption{S-mode in the SDF phase}
\label{s-mode}
\efig 

Instabilities which induce a phase crossover to vortex are expected to be caused typically by excitations of the C-mode. 
The S-mode can also induce a transition to a two vortices state (see Fig.\ref{s-mode}). 
When the exchange coupling is lowered, the population of the C-mode excited state increases as 
this mode softens and becomes finally unstable, i.e. of zero frequency. 
Near the crossover point, this mode will induce the entrance of the vortex core into the dot, without invoking an out of 
plane motion of the spins. It can be thought as the projection of a virtual supervortex with its core oscillating outside 
the dot, from near the dot boundary to near infinity.  
The lowest frequency modes in both phases are plotted as a function of 
$J$ for a circular 
96 spin dot in Fig.\ref{instability}. 
To obtain them, both the vortex and SDF structures were put as initial configurations, then relaxedi to produce the ground state configuration, 
before doing the diagonalization calculation.
We can see that 
the softening of the two phases occurs at two different exchange couplings, 
indicating that the Vortex-SDF crossover is the
analog of a "first order" transition, defined only for an infinite system. 
We believe it is caused by magnon instability.
In a transition where the two minima in the free energy are separated by a barrier, the phase change can occur by 
tunneling at zero temperature or by thermal activation at nonzero temperatures, even if $J$ is in the region where 
the C-mode frequency is still positive. This transition across the barrier occurs after a finite
time as the system itself is finite in size. At lower $J$'s this frequency becomes zero or negative. This is the region 
where there is spontaneous phase change, and where the second derivative of the energy at the
SDF spin configuration has a sign change. Therefore the crossover can take place in principle in a wide
range of $J$'s with a rate which increases as $J$ tends to the magnon instability point.

\bfig[h]
\includegraphics[angle=270, width=7 cm]{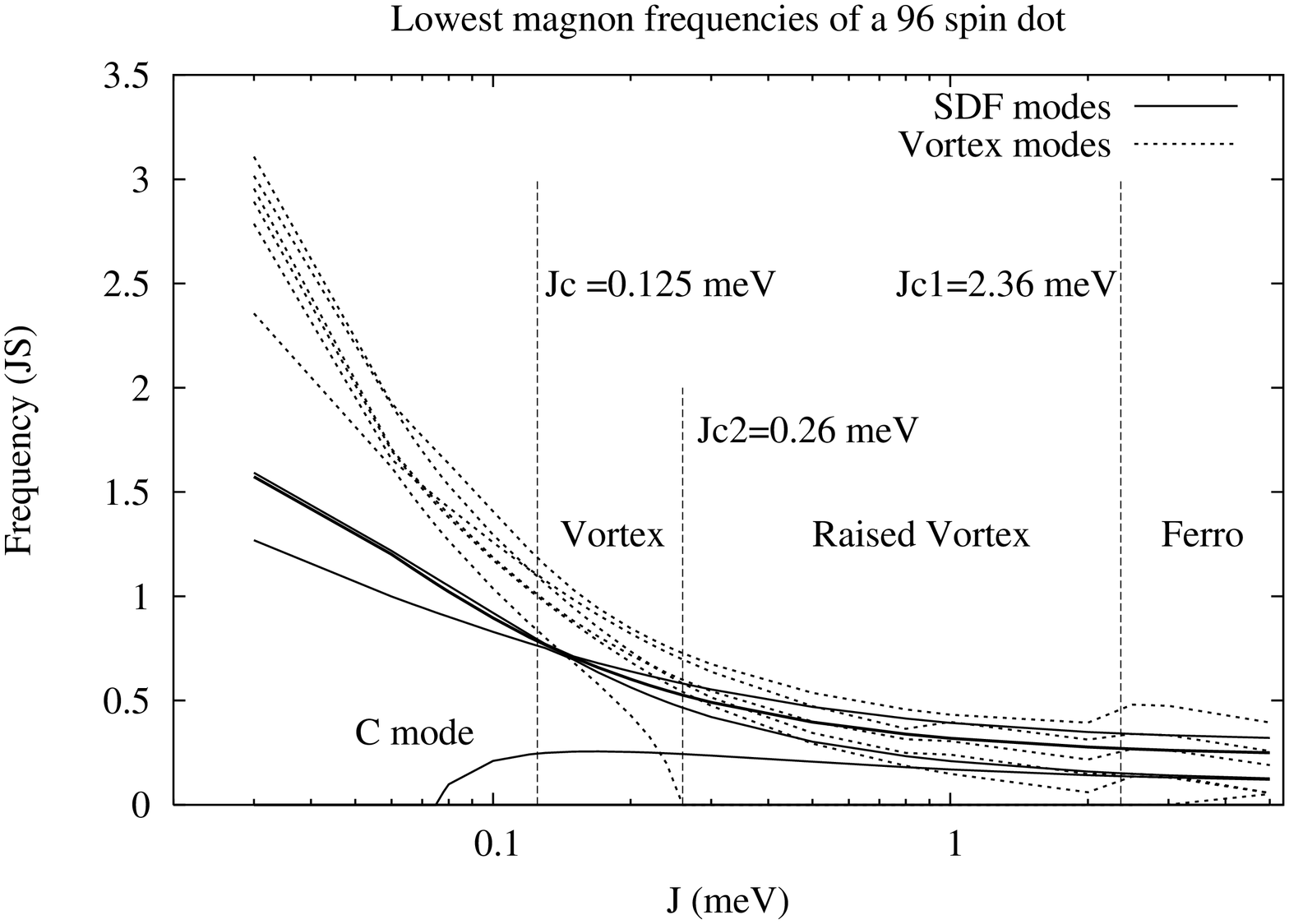}
\caption{Lowest magnon frequencies as a function of the exchange coupling 
near the transition point in both phases for a 96-spin dot. 
The critical $J$ where the total energies of these phases become equal is 
also shown ($J_c=0.125 $ meV). Magnon instability of each phase, however, 
occurs at a further point ($J=0.075$ meV : C-mode instability of SDF, 
$J=0.26$ meV : vortex raising, and $J=2.36$ meV : spontaneous crossover
to SDF). In principle, the crossover to the other phase can take place 
after a finite time for any $J$ between 0.075 and 2.36 meV .}
\label{instability}
\efig

\subsubsection{Vortex modes:}

Similar to the work of Ivanov\cite{ivanov}, we have observed the in-plane 
oscillations of the 
vortex center and shape as two of the lowest magnon modes. 
These two modes switch in order as the exchange 
coupling $J$ is increased away from its critical value. 
Unlike their prediction, however, we have seen that 
near the transition ($J \to 0.26 meV$), yet another phase appears and 
a different mode with the lowest frequency causes instability. 
This mode describes in-phase, out of plane oscillations of 
the core spins causing the instability towards a vortex with an 
out of plane magnetization localized at its core. 
As the coupling is increased, in order to reduce the frustration of the 4 
spins at the core, the vortex core acquires a finite magnetization 
perpendicular to the plane, and it also weakly oscillates 
about the center of the dot. 
This softest mode is displayed in Fig. \ref{softest-vortex} where the 
viewing direction has been slightly tilted in
order to better see out of plane spins.
Calculations which somehow confined the motion of spins in 
the plane of the disk, did not predict this lowest frequency mode. 
The order of the modes depends on the value of $J$. The order we are 
reporting here, is obtained near the magnon instability points, i.e. $J=0.25 meV < J_c = 0.26 meV$
for the vortex phase and $J=0.08 meV > J_c=0.075 meV$ for the SDF phase. Note that the total energy of 
these phases become equal at $J=0.125 meV$.

During the crossover, there is a change in the order
parameters of the system. If we define the latter by 
\beqnar
{\vec V} &=& {1 \over N} \sum_{i=1}^N \, {\ri \times \si \over ||\ri|| \, ||\si||} \\
{\vec M} &=& {1 \over N} \sum_{i=1}^N \, { \si \over ||\si||},
\label{op}
\eeqnar
then in the vortex phase we had ${\vec V} =(0,0,1)$ and ${\vec M} =0$, and in the new phase
${\vec V} =(0,0,v)$ and ${\vec M} =(0,0,m)$ where $0<v<1$ and $0<m<1$ are two real numbers.
For the considered monolayer dot, this phase is higher in energy than 
the SDF phase (${\vec V}=0$ and ${\vec M} =(m_x,m_y,0)$) and is only metastable.
It could, however, become more stable than the SDF if the number of 
layers is increased with the radius of the dot kept constant. 

Thus, it seems that although this phase is higher in energy than the 
SDF phase, the system goes from the in-plane vortex, to this one which 
we call a raised vortex, then by tunneling, or if $J$ becomes large 
enough, directly, to a ferromagnetic phase.
For this sample, we observed that if $J$ becomes as large as $2.36 meV$, 
$v$ drops to zero and $m$ is substantially increased ($m < 1$), 
meaning that the vortex core, which is ferromagnetic, is enlarged 
till encompassing the whole dot. This intermediate metastable phase 
becomes unstable at 2.36 and the
magnetization spontaneously lies in the plane for larger exchange couplings.  

\bfig[h]
\includegraphics[width=7 cm]{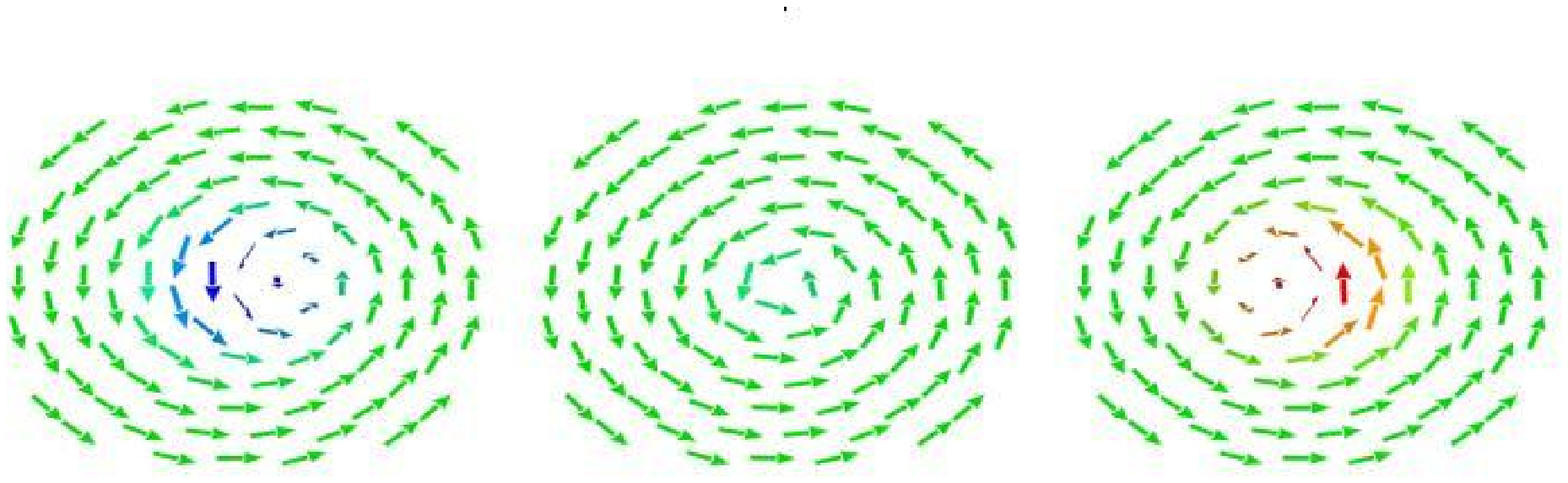}
\caption{Softest mode in vortex phase near the critical point ($J=0.25$ meV). The plane of the 
disk is tilted in order to show the out of plane component of the core spins. 
One can identify an oscillation of the out of plane component of the 
core spins.}
\label{softest-vortex}
\efig 
The second mode is displayed in Fig. \ref{sawtooth}. We have called it 
the sawtooth mode as all spins around a ring oscillate in phase just as a 
sawtooth.  This mode, we believe has not been reported in the past. 
It has no nodal lines, confirming its low frequency, and is the curled 
up version of the Goldstone mode of the SDF phase.

\bfig[h]
\includegraphics[width=7 cm]{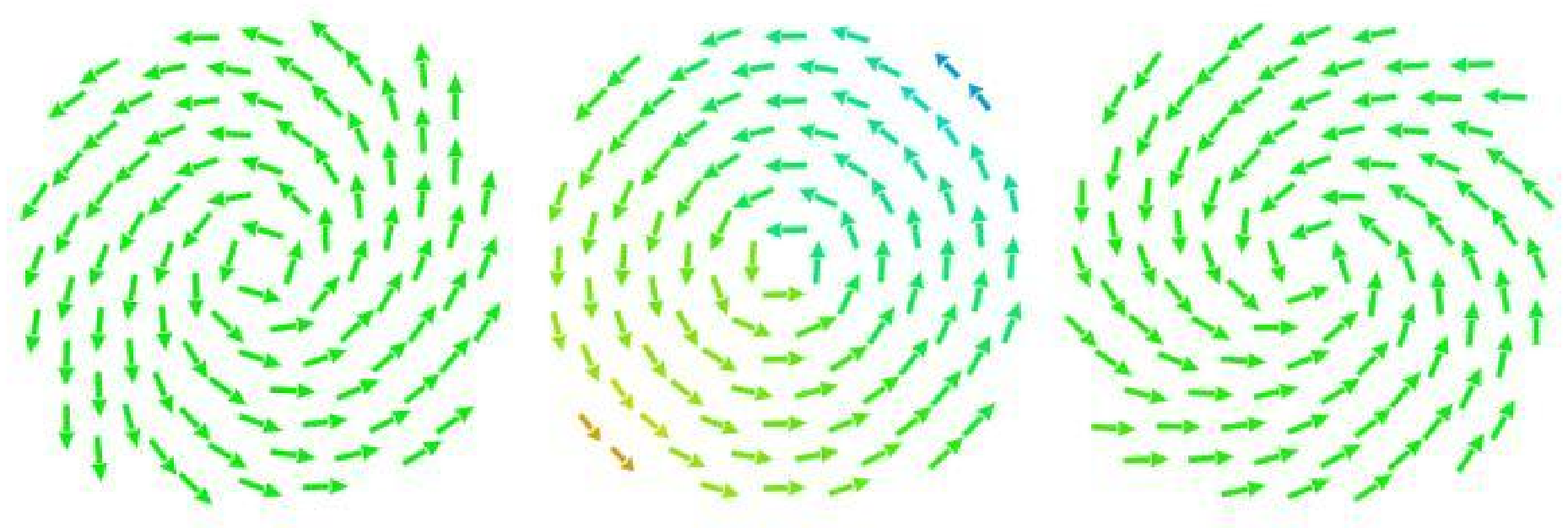}
\caption{In the saw-tooth mode, which has no nodal line, spins around a ring perform an in-phase oscillation in the plane of the dot. }
\label{sawtooth}
\efig 

The next two lowest modes, which are well-known, are the oscillations of the vortex core and of its shape.
They are displayed in Figs. \ref{core-oscillation} and \ref{core-shape}.
It was expected that the oscillations of the core center would cause 
the instability to the SDF phase: 
as $J$ is increased, this mode would soften, making the amplitude of the 
core oscillations larger, until the center is kicked out of 
the sample and one ends up with a single ferromagnetic domain. 
Our results on the magnon instability, however, show otherwise: 
as $J$ is increased, first the core spins raise out of the plane and the 
modes are more or less similar to those of the in-plane vortex. 
So the ground state, as we previously described, has $0<m<1$ and $0<v<1$ with
$M$ increasing and $v$ decreasing as $J$ is increased. For instance, near 
$J \simeq 2$ meV, the
magnetization is more localized at the center, and the lowest mode consists
in oscillations of the vortex core combined with precession of spins about
their ground state value (the second mode being still the sawtooth). Although
ultimately the lowest mode has core oscillations, at the same time the spins
are lined up perpendicular to the plane of the sample, and the transition to
the ferromagnetic state includes both the departue of the core from the sample
and the switching of the magnetization to the in-plane direction due 
to the dipole-induced anisotropy.

\bfig[h]
\includegraphics[width=7 cm]{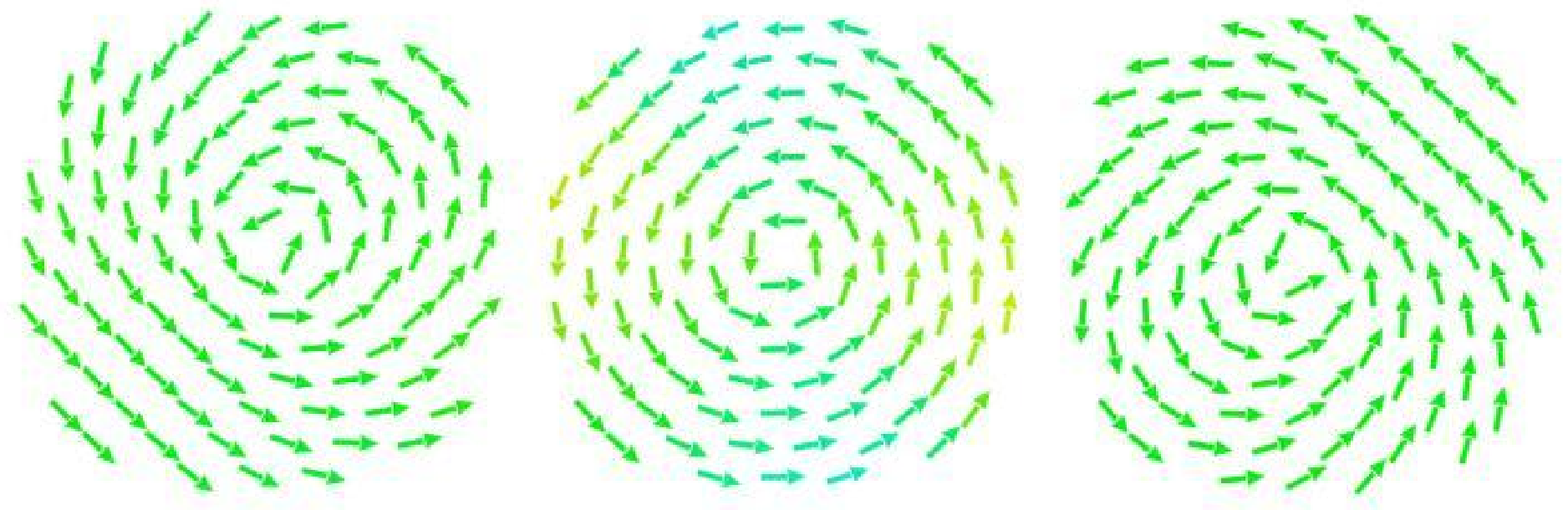}
\caption{Oscillations of the vortex core near the transition.}
\label{core-oscillation}
\efig 
\bfig[h]
\includegraphics[width=7 cm]{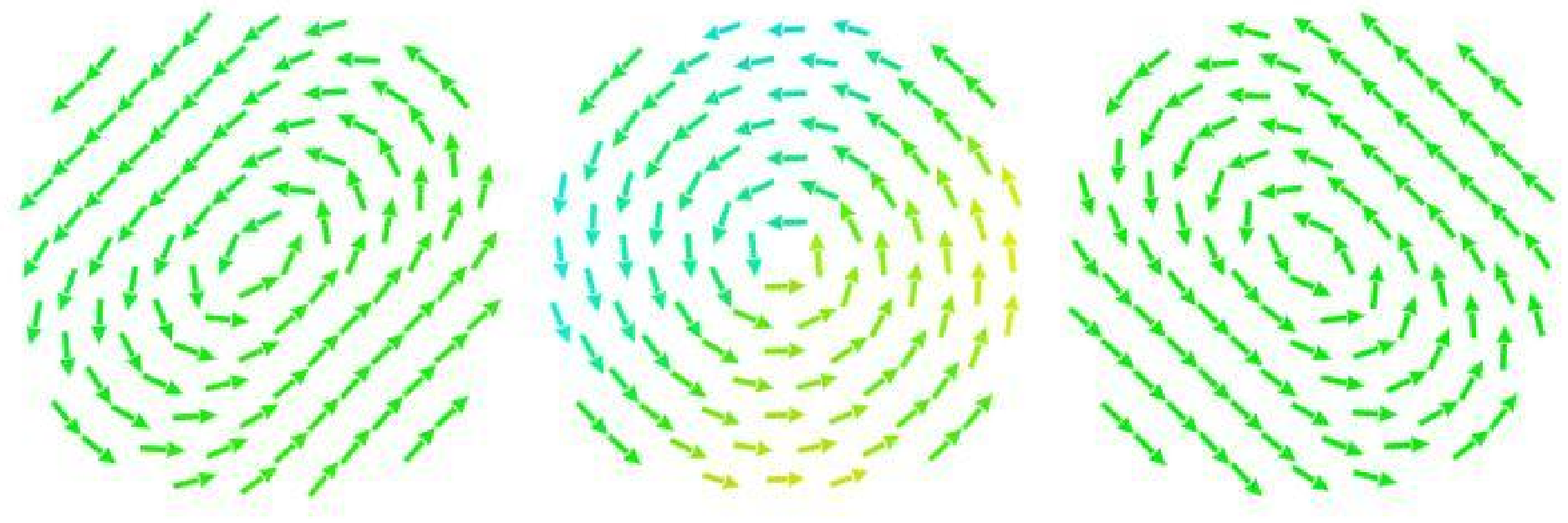}
\caption{Oscillations of the core shape. During the oscillations, the spins remain in the plane.}
\label{core-shape}
\efig 

In higher frequency modes, large amplitude in plane oscillations of the spins of central rings are observed; 
whereas spins at the dot center are less mobile. 
As frequencies become yet higher, one can see shorter wavelength 
(with 2 or mode nodal lines along the circumference) 
Larmor-like precessions of the spins around the vortex rings. Still other modes consist in out of plane oscillations 
of the spins with wavelengths varying from two lattice spacing at high frequencies, to the vortex circumference 
at lower frequencies. Yet another class of modes have nodal lines in the radial direction. 
In high frequency modes, nodal lines are both radial and along the circumference.

In recent experiments on detecting magnon frequencies\cite{park}, very few 
modes have been observed. In principle, there are as many modes as there 
are spins in the system. Presumably very specific modes are excited by 
the pulse in the experiment. Furthermore, modes of frequency lower than 
the relaxation rate associated with the Gilbert damping term, are never 
observed since before any oscillation occurs they are damped. However, 
modes such as vortex core oscillations that we have identified here, 
have been observed.

\section{Conclusions}

To summarize, the energetics and dynamics of semiclassical spins interacting via exchange and dipole fields was 
considered in this work. Two phases were identified and their total energy was formulated in the continuum approximation. 
The crossover was investigated by comparison of total energies and its mechanism described by magnon instability.
Magnon frequencies in each phase were calculated and characterized for a finite size disk-shape dot.
A metastable intermediate phase was also identified for the monolayer $N=96$ spin system. This could become a
stable phase if more than one layer are involved. The in-plane vortex phase first goes through
this ``raised core" phase where the sample acquires an out of plane magnetization in the core region
of the vortex. Then as $J$ is further increased, the latter tunnels to the in-plane ferromagnetic phase.

Magnon modes of the SDF phase, neglecting the boundary effects, consist in spin precessions about the 
equilibrium value where the motion is either in plane or out of plane or eventually mixed. The precessions 
of neighboring spins are different by a phase which is $\pi$ for high frequency modes and nearly zero ($\pi a/L$) for 
low frequency ones. Nodal lines are perpendicular to each other and increase in number as the frequencies go up,
although this is not true for small systems where boundary effects are important. 
In the vortex phase, the nodal lines are radial and also along the circumference. A low frequency mode
which we called ``saw tooth" was identified, and not yet reported to the best of our knowledge. It
is however believed that the transition to the SDF phase takes place via the oscillations of the 
vortex core.

The finite temperature behavior of the vorticity and magnetization was described in terms of magnon
density of states. 

\section{Acknowledgements}
The authors would like to thank the ICTP for their hospitality where a part of this work was done.
MRM would like to acknowledge discussions with S. M. Fazeli and S. Abedinpour.

%\newpage

%\end{references}


\begin{thebibliography}{99}
\bibitem{goldstone}
The frequency is not strictly zero as there is a very small in-plane anisotropy.
\end{thebibliography}
\end{document}